\documentclass[12pt, a4paper]{article}
\usepackage{cite}
\usepackage{amsmath,amssymb}
\input{colordvi.tex}
\usepackage{comment}
\usepackage{bm}
\usepackage{url}

\usepackage{ifpdf}
\ifpdf        
  \usepackage{graphicx, hyperref, xcolor}     
\else     
  \usepackage[dvipdfmx]{graphicx, hyperref, xcolor}     
 \fi

\definecolor{rossoferrari}{HTML}{D9073D}
\definecolor{mediumblue}{HTML}{0000CD}
\hypersetup{
setpagesize=false,
bookmarksnumbered=true,%
bookmarksopen=true,%
colorlinks=true,%
linkcolor=rossoferrari,
urlcolor=mediumblue,
citecolor=mediumblue,
}

\begin{document}

\begin{titlepage}

\begin{center}

\hfill UT-20-04\\

\vskip .75in

{\Large \bf 
Constraint on Vector Coherent Oscillation \\[.3em] Dark Matter with Kinetic Function
}

\vskip .75in

{\large
Kazunori Nakayama$^{(a,b)}$
}

\vskip 0.25in

$^{(a)}${\em Department of Physics, Faculty of Science,\\
The University of Tokyo,  Bunkyo-ku, Tokyo 113-0033, Japan}\\[.3em]
$^{(b)}${\em Kavli IPMU (WPI), The University of Tokyo,  Kashiwa, Chiba 277-8583, Japan}

\end{center}
\vskip .5in

\begin{abstract}

A spatially uniform vector condensate can be formed during inflation if the vector boson is coupled to the inflaton through nontrivial kinetic function. The coherent oscillation of such a massive vector boson is a dark matter candidate. In this paper we consider the case where the vector boson energy density increases during inflation and show that the curvature/isocurvature perturbation gives stringent constraint on this scenario.

\end{abstract}

\end{titlepage}


\renewcommand{\thepage}{\arabic{page}}
\setcounter{page}{1}
\renewcommand{\thefootnote}{\#\arabic{footnote}}
\setcounter{footnote}{0}

\newpage

\tableofcontents

\section{Introduction}
\label{sec:Intro}

Very light bosonic dark matter (DM) scenarios recently draw lots of attention~\cite{Jaeckel:2010ni,Arias:2012az}. Axion-like particle is the most widely studied scenario in this class of models, but a massive vector boson is also a plausible DM candidate. One caveat of the vector DM model is that it is a bit nontrivial to obtain a correct abundance of the present DM compared with the case of scalar field.

So far several mechanisms to produce a correct amount of vector DM have been proposed: vector boson production through the (pseudo-)scalar coupling~\cite{Agrawal:2018vin,Co:2018lka,Bastero-Gil:2018uel,Dror:2018pdh}, inflationary fluctuation~\cite{Graham:2015rva}, gravitational particle production~\cite{Ema:2019yrd}, production through the cosmic string dynamics~\cite{Long:2019lwl} and coherent oscillation of the vector boson~\cite{Nelson:2011sf,Arias:2012az,AlonsoAlvarez:2019cgw,Nakayama:2019rhg}.

In this paper we focus on the coherent oscillation scenario of vector boson DM. 
Ref.~\cite{Nelson:2011sf} considered a minimal massive vector field, but actually the vector energy density exponentially damps during inflation and this minimal model does not work. Refs.~\cite{Arias:2012az,AlonsoAlvarez:2019cgw} considered a non-minimal vector boson coupling to the Ricci curvature in order to sustain a vector condensate during inflation, but this introduces a ghost instability of the longitudinal fluctuation~\cite{Dvali:2007ks,Himmetoglu:2008zp,Himmetoglu:2009qi,Karciauskas:2010as}, as pointed out in Ref.~\cite{Nakayama:2019rhg}.
A possible extension without such an instability is to introduce a kinetic coupling to the inflaton $\phi$ through the form of $\mathcal L \sim f^2(\phi) F_{\mu\nu}F^{\mu\nu}$~\cite{Nakayama:2019rhg}. If the kinetic function $f(\phi)$ has a particular time dependence, the vector boson condensate does not decay during inflation and one can sustain a homogeneous vector field. Later it begins a coherent oscillation and it behaves as non-relativistic matter. Such a scenario was extensively studied in the context of vector curvaton~\cite{Dimopoulos:2007zb,Dimopoulos:2009am,Dimopoulos:2009vu,Dimopoulos:2010xq}.
In Ref.~\cite{Nakayama:2019rhg} the case of $f^2(\phi) \propto a(t)^{\alpha}$ with $\alpha=-4$ or $2$, where $a(t)$ denotes the cosmic scale factor, was considered as an illustration and it was shown that the physical vector field value\footnote{
	The ``physical'' vector field is defined later below Eq.~(\ref{rhoA}).
} can remain constant during inflation for this particular choice. However, there is a priori no reason to choose this parameter and parameter tuning is required for this scenario to realize. 

In this paper we mainly study the case of $\alpha<-4$ or $\alpha>2$. In this case the physical vector boson field is amplified during inflation. The vector boson energy density grows and eventually the backreaction of the vector field to the inflaton dynamics becomes important. It is known that in such a case the background vector boson can support another inflation regime, called the anisotropic inflation~\cite{Watanabe:2009ct,Soda:2012zm,Maleknejad:2012fw}. The vector boson energy density is saturated at some value during the anisotropic inflation and we will consider a possibility that this vector boson will later become a coherent oscillation DM. 

In Sec.~\ref{sec:dyn} we study the dynamics of the inflaton and vector boson of the homogeneous mode. It is found that the vector boson in this scenario can indeed have a correct abundance as total DM. In Sec.~\ref{sec:pert} we discuss constraints on this scenario through the properties of the curvature and isocurvature perturbation. Actually they give very stringent constraint and we are left with only a limited possibility as a consistent DM scenario. Sec.~\ref{sec:dis} is devoted to conclusions and discussion.

\section{Dynamics of vector condensate} \label{sec:dyn}

We consider the following action for a massive vector boson $\mathcal A_M$ and inflaton $\phi$:
\begin{align}
	S = \int d^3x dt\sqrt{-g} \left[-\frac{f^2(\phi)}{4} \mathcal F_{MN}\mathcal F^{MN} -\frac{h^2(\phi)}{2}m_A^2 \mathcal A_M \mathcal A^M -\frac{1}{2} \partial_M \phi\,\partial^M \phi -V(\phi)\right].
	\label{action}
\end{align}
We have introduced kinetic function $f(\phi)$ and mass function $h(\phi)$, whose functional form will be given later. Taking account of the effect of the background homogeneous vector field, whose direction is taken to be $x$ direction without loss of generality, i.e. $\mathcal A_i=(A,0,0)$, the metric is taken to be the so-called Bianchi type-I form:
\begin{align}
	ds^2 = -dt^2 + a^2(t) \left[ e^{-4\sigma(t)} dx^2 +e^{2\sigma(t)}(dy^2+dz^2) \right],
\end{align}
where $a(t)$ is the cosmic scale factor and $\sigma(t)$ represents the anisotropic expansion.
For a while we neglect the anisotropy, i.e. $\sigma=0$, by assuming that the energy density of the vector field is much smaller than the inflaton. This will be justified later.

It is often useful to rescale the vector field as $A_\mu = (a\mathcal A_0,\mathcal A_i)$ and use the conformal time $d\tau = dt/a$ to obtain the action
\begin{align}
	S=\int d\tau d^3x \left[ -\frac{f^2}{4} \eta^{\mu\rho}\eta^{\nu\sigma}F_{\mu\nu}F_{\rho\sigma}-\frac{a^2 h^2}{2}m_A^2 \eta^{\mu\nu} A_\mu A_\nu \right].
\end{align}
Below we consider the special form of the kinetic function:
\begin{align}
	f(\phi) = \exp\left( -\frac{\gamma}{2M_P^2}\int\frac{V}{V_\phi}d\phi \right),
\end{align}
where $V_\phi \equiv \partial V/\partial\phi$. A particular example is the chaotic inflation:
\begin{align}
	f(\phi) = \exp\left( -\frac{\gamma}{4n}\frac{\phi^2}{M_P^2} \right),~~~~~~V(\phi)=\frac{\lambda \phi^n}{n}.   \label{chaotic}
\end{align}
For the new (hilltop) inflation model, it is given by
\begin{align}
	f(\phi) = \exp\left( -\frac{\gamma}{4n(n-2)}\frac{v^n}{M_P^2 \phi^{n-2}} \right),~~~~~~V(\phi)=\Lambda^4\left[1-\left(\frac{\phi}{v}\right)^n\right]^2.
	\label{new}
\end{align}
They lead to the scaling of $f^2 \propto a^\gamma$ during the standard slow-roll inflation when the effect of backreaction of the vector field to the inflaton dynamics is negligible. Note also that $f\simeq 1$ soon after inflation ends. The case of $\gamma=-4$ and $2$ have been discussed in Ref.~\cite{Nakayama:2019rhg}. We will consider the case of $\gamma < -4$ and $\gamma>2$.
For the moment we do not assume any specific functional form of $h(\phi)$ except that it soon approaches to $h(\phi)\to 1$ after inflation ends.

\subsection{Dynamics during inflation} \label{sec:inf}

Let us describe the vector and inflaton dynamics during inflation neglecting the spatial fluctuation. We follow the analysis given in Refs.~\cite{Watanabe:2009ct,Soda:2012zm,Maleknejad:2012fw}. The equation of motion is given by
\begin{align}
	&a\frac{\partial}{\partial t}\left( af^2 \dot A \right) + a^2 h^2 m_A^2 A = 0, \label{eom_A}\\
	&\ddot\phi + 3H \dot\phi + V_\phi -\frac{1}{a^2}f f_\phi \dot A^2 = 0,   \label{eom_p}
\end{align}
where $f_\phi \equiv \partial f/\partial\phi$. The Hubble parameter $H= \dot a/a$ is given by
\begin{align}
	3M_P^2 H^2 = \rho_\phi + \rho_A,
\end{align}
with $M_P$ being the reduced Planck scale and $\rho_\phi$ the inflaton energy density. The vector boson energy density $\rho_A$ is given by
\begin{align}
	\rho_A = \frac{1}{2a^2}\left(f^2\dot A^2 + h^2 m_A^2 A^2 \right)
	= \frac{1}{2}\left\{ \left[\dot{\overline{A}} +\left(H-\frac{\dot f}{f}\right) \overline{A} \right]^2 + \frac{h^2 m_A^2}{f^2}\overline{A}^2 \right\},
	\label{rhoA}
\end{align}
where we have defined the ``physical'' vector field as $\overline{A} \equiv fA/a$~\cite{Nakayama:2019rhg}.

Let us consider the case where the vector boson mass term can be safely ignored. Then we immediately obtain 
\begin{align}
	a f^2 \dot A = {\rm const.}
\end{align}
Supposing the scaling $f^2\propto a^\alpha$ with $\alpha$ being a numerical constant, we schematically obtain
\begin{align}
	A = C_1+ C_2 a^{-(1+\alpha)},   \label{A_inf}
\end{align}
during inflation where $C_1$ and $C_2$ are constants. As soon explained below, $\alpha=\gamma$ when the vector energy density is negligible but $\alpha$ can take different value from $\gamma$ when the backreaction is important. The physical field roughly behaves as
\begin{align}
	\overline A \propto 
	a^{\left(|1+\alpha|-3 \right)/2} =
	\begin{cases}
		a^{\alpha/2-1} & {\rm for}~~1+\alpha>0\\
		a^{-\alpha/2-2} & {\rm for}~~1+\alpha\leq0
	\end{cases}.
	\label{barAscaling}
\end{align}
Note that, although $\overline A$ is increasing for $\alpha<-4$ and $\alpha>2$, the $C_1$ term does not contribute to the kinetic energy in (\ref{rhoA}). Since the $C_1$ term is dominant for $1+\alpha>0$, actually the energy density is actually decreasing for $\alpha>2$.
In both cases the vector energy density scales as $\rho_A \propto f^{-2} a^{-4} \propto a^{-\alpha-4}$. 
Below we consider the case of $\gamma <-4$ and $\gamma>2$ separately.

\subsubsection{$\gamma < -4$}

As studied in Ref.~\cite{Nakayama:2019rhg}, $\rho_A$ remains constant for $\gamma=-4$ that ensures the establishment of the vector boson homogeneous condensate during inflation.
On the other hand, it increases during inflation for $\gamma< -4$ and hence eventually the backreaction will become important.
The inflaton equation of motion is written as
\begin{align}
	 \ddot\phi + 3H \dot\phi + V_\phi\left( 1 + \frac{\gamma}{2\epsilon_V} \frac{\rho_A}{V} \right) = 0,  \label{eom_phi}
\end{align}
where $\epsilon_V \equiv M_P^2(V_\phi/V)^2/2$ is the slow-roll parameter.\footnote{
	Note that $\epsilon_H \equiv -\dot H/H^2 = -(4/\gamma)\epsilon_V$.
}
Thus it is seen that if the vector boson energy density satisfies $\rho_A \ll (2\epsilon_V/|\gamma|)V$, the effect of the vector boson on the inflaton dynamics is safely neglected. 
For $\gamma< -4$, however, $\rho_A$ increases during inflation and $\rho_A$ will become comparable to $(2\epsilon_V/|\gamma|)V$.
It is expected that $\phi$ will slow down at this stage since the parenthesis in the last term of Eq.~(\ref{eom_phi}) will approach to zero, which effectively ``flattens'' the inflaton potential. Correspondingly the time evolution of the function $f(\phi)$ also changes so that $\rho_A$  approximately remains constant: $f^{-2} a^{-4} \sim {\rm const}.$ This requires the following relation:
\begin{align}
	\dot\phi \simeq \frac{4H M_P^2 V_\phi}{\gamma V} = \frac{4}{3\gamma H} V_\phi.
\end{align}
In order for this solution to be consistent with slow-roll equation of (\ref{eom_phi}), the energy density should satisfy
\begin{align}
	\frac{\rho_A}{\rho_\phi} = -2\epsilon_V\frac{\gamma+4}{\gamma^2} \equiv R_A.   \label{RA}
\end{align}
Here $\rho_\phi\simeq V$ is the inflaton energy density. To summarize, the slow-roll inflaton dynamics is described by
\begin{align}
	3H \dot \phi \simeq \begin{cases}
		\displaystyle -V_\phi &\displaystyle {\rm for}~~~\frac{\rho_A}{\rho_\phi} \ll R_A,\\
		\displaystyle \frac{4}{\gamma}V_\phi &\displaystyle {\rm for}~~~\frac{\rho_A}{\rho_\phi} \simeq R_A\\
	\end{cases}.
\end{align}
One can see that the potential is effectively flattened by a factor $-4/\gamma$ due to the vector backreaction.
This second case is a slow-roll inflation supported by the vector field and it is called the anisotropic inflation because the vector field condensate implies a preferred direction. Even if the initial vector energy density is negligibly small, it will be exponentially amplified during inflation and it enters the regime of anisotropic inflation, although still the vector energy density is much smaller than the inflaton itself at this stage.

\begin{figure}
\begin{center}
\begin{tabular}{cc}
\includegraphics[scale=1.3]{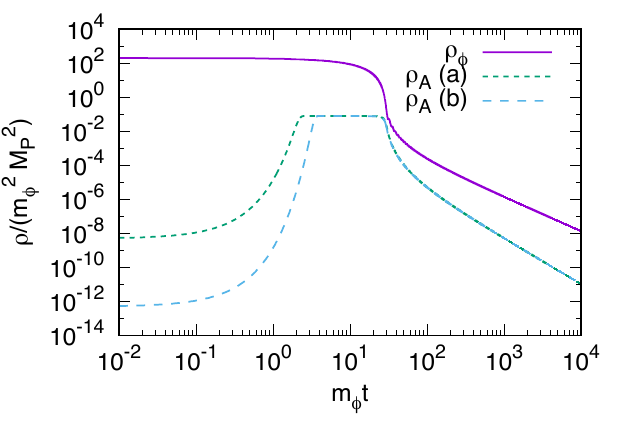}
\includegraphics[scale=1.3]{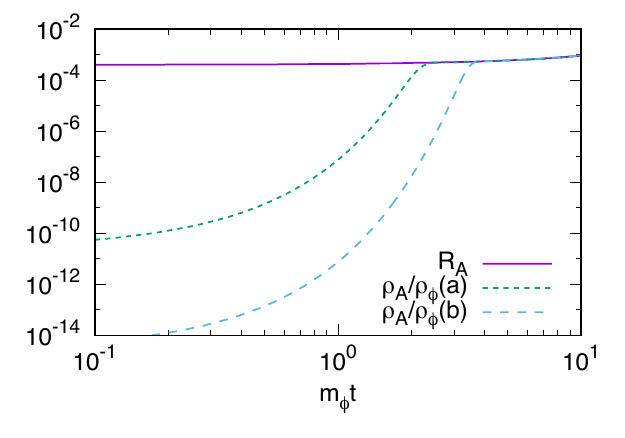}
\end{tabular}
\end{center}
\caption{
	Numerical results for chaotic inflation model (\ref{chaotic}) with $n=2$. (Left) Time evolution of the energy density of the inflaton $(\rho_\phi)$ and vector boson $(\rho_A)$ normalized by $m_\phi^2 M_P^2$. We have taken $\gamma=-5$ and $\dot{\overline A}=10^{-4} m_\phi M_P$ for (a) and $10^{-6} m_\phi M_P$ for (b) as initial condition. (Right) Time evolution of the ratio $\rho_A/\rho_\phi$ compared with $R_A$. Parameters are the same as the left panel.
}
\label{fig:chao}
\end{figure}

\begin{figure}
\begin{center}
\begin{tabular}{cc}
\includegraphics[scale=1.3]{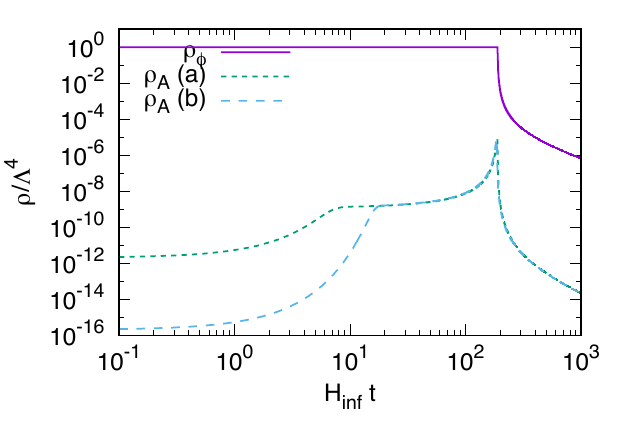}
\includegraphics[scale=1.3]{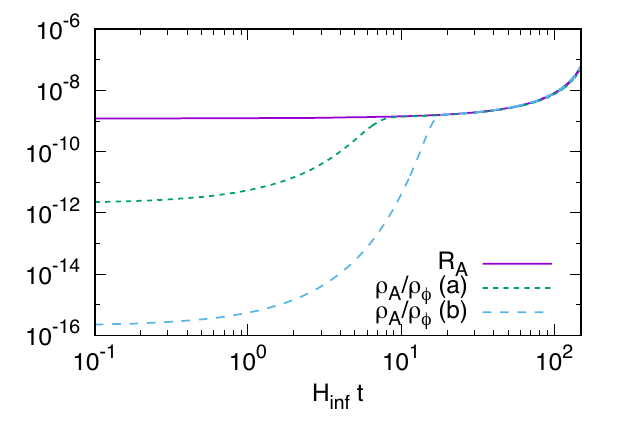}
\end{tabular}
\end{center}
\caption{
	Numerical results for new inflation model (\ref{new}) with $n=6$. (Left) Time evolution of the energy density of the inflaton $(\rho_\phi)$ and vector boson $(\rho_A)$ normalized by $\Lambda^4$. We have taken $\gamma=-5$ and ${\overline A}=10^{-6}M_P$ for (a) and $10^{-8}M_P$ for (b) as initial condition. (Right) Time evolution of the ratio $\rho_A/\rho_\phi$ compared with $R_A$. Parameters are the same as the left panel.
}
\label{fig:new}
\end{figure}

Fig.~\ref{fig:chao} shows the result of numerical solution of the equation of motion of the inflaton (\ref{eom_p}) and vector boson (\ref{eom_A}) for the inflaton potential $V=m_\phi^2 \phi^2/2$ and $\gamma=-5$. Time evolution of the energy density of the inflaton $(\rho_\phi)$ and vector boson $(\rho_A)$ normalized by $m_\phi^2 M_P^2$ are shown in the left panel. We have taken $\phi=20M_P$ and $\dot{\overline A}=10^{-4} m_\phi M_P$ for (a) and $10^{-6} m_\phi M_P$ for (b) as initial conditions and the massless limit $m_A\to 0$. Time evolution of the ratio $\rho_A/\rho_\phi$ compared with $R_A$ (\ref{RA}) is shown in the right panel. Parameters are the same as the left panel. 
Similarly, Fig.~\ref{fig:new} shows the result of numerical calculation for the new inflation model (\ref{new}) with $n=6$. We have taken $\gamma=-5$, $v=M_P$, $\phi=0.3\phi_{\rm end}$ and ${\overline A}=10^{-6}M_P$ for (a) and $10^{-8}M_P$ for (b) as initial condition. Here $\phi_{\rm end}$ denotes the inflaton field value at which inflation ends: $2n(n-1)(\phi_{\rm end}/v)^{n-2}= v^2/M_P^2$.
In both cases it is clearly seen that the ratio $\rho_A/\rho_\phi$ approaches to the value given by $R_A$ (\ref{RA}) independently of the initial condition.

Note that in the calculation performed for these figures the anisotropic inflation regime does not last for very long time, but it is an artifact of the choice of the initial condition. If, for example, the calculation starts from much larger (smaller) inflaton field value for the chaotic (new) inflation model, the vector boson density is saturated at much earlier time and the anisotropic inflation lasts for much longer time (say, much longer than 60 e-folds).
We do not go into details of the problem of initial condition since it is related with the dynamics before the ``observable'' inflation happens and just treat the initial condition as free parameters.\footnote{
	It is possible that the long wavelength vector perturbation accumulates to constitute a ``homogeneous'' mode if the total duration of inflation is long enough~\cite{Bartolo:2012sd,Sanchez:2013zst}.
}

So far we have ignored the anisotropic expansion. The equation for $\Sigma \equiv \dot \sigma$ is given by
\begin{align}
	\dot\Sigma + 3H\Sigma = \frac{2\rho_A}{3M_P^2}.
\end{align}
It is expected that $\Sigma$ converges to a nearly constant value
\begin{align}
	\frac{\Sigma}{H} \simeq \frac{2\rho_A}{3\rho_\phi}= -\frac{4\epsilon_V}{3}\frac{\gamma+4}{\gamma^2} = \frac{\epsilon_H}{3} \frac{\gamma+4}{\gamma}.
\end{align}
It is suppressed by the slow-roll parameter $\epsilon_H$. Therefore the homogeneous dynamics is not much affected by the inclusion of the anisotropic expansion.

\subsubsection{$\gamma > 2$}   \label{sec:inf2}

In this case, as far as the vector boson mass $m_A$ is negligible, the vector energy density $\rho_A$ (or its kinetic part) decreases as $\rho_A\propto a^{-\gamma-2}$ during inflation and hence it rapidly approaches to zero, while the $\overline A$ increases as $\overline A \propto a^{\gamma/2-1}$. Since $\rho_A$ is negligible, there is no backreaction of the vector field to the inflaton and the anisotropic inflation does not occur.

On the other hand, the condition that the vector boson mass is negligible is written as $h m_A/f \ll H_{\rm inf}$ at least during the last 60 e-foldings of inflation. For $h=1$ for example, it gives a constraint on the vector boson mass as
\begin{align}
	m_A \ll e^{-30\gamma} H_{\rm inf} \sim 10^{-13 \gamma}H_{\rm inf}.
\end{align}
Then the vector boson energy density at the end of inflation is bounded as
\begin{align}
	\left.\frac{\rho_A}{\rho_\phi}\right|_{\tau_{\rm end}} \ll 10^{-26\gamma}.   \label{rhoA_bound}
\end{align}
If the total duration of inflation is much longer than 60 e-foldings, the constraint becomes much more stringent. If this condition is violated, the vector boson mass would make rapid decay of the amplitude $\overline A$ during inflation.

Another choice is $h(\phi)=f(\phi)$. In this case, the only requirement is $m_A \ll H_{\rm inf}$. Thus the upper bound on the vector energy density at the end of inflation is just 
\begin{align}
	\left.\frac{\rho_A}{\rho_\phi}\right|_{\tau_{\rm end}}\simeq \frac{m_A^2 \overline A^2}{6H_{\rm inf}^2 M_P^2} \ll \left( \frac{\overline A}{M_P} \right)^2.
\end{align}

\begin{figure}
\begin{center}
\begin{tabular}{cc}
\includegraphics[scale=1.3]{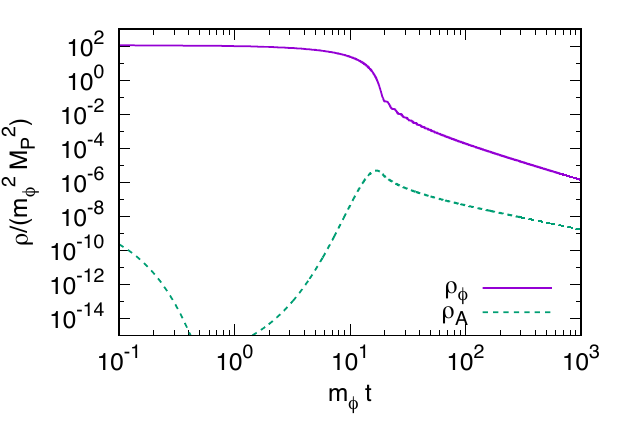}
\includegraphics[scale=1.3]{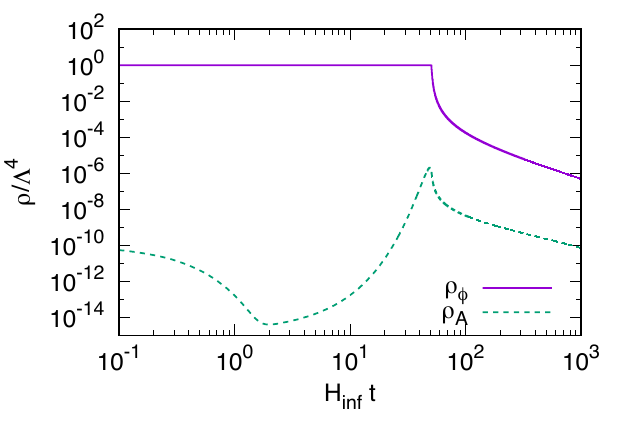}
\end{tabular}
\end{center}
\caption{
	Time evolution of the energy density of the inflaton $(\rho_\phi)$ and vector boson $(\rho_A)$ for $\gamma=5/2$ and $h(\phi)=f(\phi)$ for chaotic inflation model  (\ref{chaotic}) with $n=2$ (left) and new inflation model  (\ref{new}) with $n=6$ (right). The vector boson mass is taken to be $m_A=10^{-5} H_{\rm inf}$ (left) and $m_A=10^{-3} H_{\rm inf}$ (right). As initial condition, we have taken ${\overline A}=10^{-4}M_P$, $\phi=15M_P$ (left) and $\phi=0.4\phi_{\rm end}$ (right).
}
\label{fig:25}
\end{figure}

Fig.~\ref{fig:25} shows the time evolution of the energy density of the inflaton $(\rho_\phi)$ and vector boson $(\rho_A)$ for $\gamma=5/2$ and $h(\phi)=f(\phi)$ for chaotic inflation model  (\ref{chaotic}) with $n=2$ (left) and new inflation model  (\ref{new}) with $n=6$ (right). The vector boson mass is taken to be $m_A=10^{-5} H_{\rm inf}$ (left) and $m_A=10^{-3} H_{\rm inf}$ (right). As initial condition, we have taken ${\overline A}=10^{-4}M_P$, $\phi=15M_P$ (left) and $\phi=0.4\phi_{\rm end}$ (right).
It is seen that first $\rho_A$ decreases exponentially but later the mass term begins to dominate and it increases until the end of inflation.

\subsection{Dynamics after inflation}

After inflation ends, the kinetic function and mass function is taken to be $f\simeq h\simeq 1$.
The inflaton coherent oscillation behaves as non-relativistic matter until the reheating is completed at $H = \Gamma_{\phi}$ where $\Gamma_\phi$ denotes the inflaton decay width. After the completion of reheating, the radiation-dominated universe begins. This thermal history is described by the equation of state parameter $w$, which takes $w=0$ $(1/3)$ for the matter (radiation)-dominated era.

The equation of motion of the vector field $\overline A \equiv fA/a$ is given by
\begin{align}
	\ddot{\overline A} + 3H \dot{\overline{A}} + \left(m_A^2 + \frac{1-3w}{2}H^2 \right) \overline A = 0.
\end{align}
For $H \gg m_A$, we find the solution to this equation as
\begin{align}
	\overline{A} = d_1 a^{-1} + d_2 a^{(3w-1)/2},  \label{Abar_after}
\end{align}
with $d_1$ and $d_2$ being some constants. Notice that the $d_1$ term corresponds to the solution $A={\rm const.}$ It is consistent with the solution during inflation for $\gamma>2$, which corresponds to the $C_1$ term in (\ref{A_inf}) during inflation. For $\gamma<-4$, on the other hand, the $d_2$ term solution applies. Below we consider the case of $\gamma<-4$ and $\gamma>2$.


\subsubsection{$\gamma <-4$}

For $\gamma<-4$, taking the $d_2$ term in (\ref{Abar_after}), we obtain $\overline{A} \propto a^{-1/2}$ for $w=0$ and $\overline{A} \propto a^{0}$ for $w=1/3$, which means $\rho_A \propto a^{-4}$ for both cases. This behavior is seen in Figs.~\ref{fig:chao} and \ref{fig:new}.
On the other hand, for $H\ll m_A$, the equation is the same as the minimal scalar field: it begins coherent oscillation at $H \sim m_A$ and behaves as non-relativistic matter thereafter. Hence we have $\rho_A \propto a^{-3}$ for $H\ll m_A$ and it is a candidate of DM.

Keeping this in mind, we can now evaluate the vector DM abundance.
First we consider the case of $\Gamma_\phi > m_A$. In this case the final energy density to the entropy density $(s)$ ratio is evaluated as
\begin{align}
	\frac{\rho_A}{s} = \left(\frac{\rho_A}{\rho_\phi}\right)_{H=\Gamma_\phi}\left(\frac{\rho_\phi}{s}\right)_{H=m_A}
	= \frac{3R_A}{4}\left( \frac{90}{\pi^2 g_*} \right)^{1/4} \left( \frac{\Gamma_\phi}{H_{\rm inf}} \right)^{2/3} \sqrt{m_A M_P},
\end{align}
where $\rho_\phi$ collectively denotes the inflaton energy density or the radiation energy density produced by the inflaton decay. For the other case $\Gamma_\phi < m_A$, we have
\begin{align}
	\frac{\rho_A}{s} = \left(\frac{\rho_A}{\rho_\phi}\right)_{H=m_A}\left(\frac{\rho_\phi}{s}\right)_{H=\Gamma_\phi}
	=  \frac{3R_A}{4} \left(\frac{m_A}{H_{\rm inf}} \right)^{2/3} T_{\rm R},
\end{align}
with $T_{\rm R}$ being the reheating temperature. Numerically they are summarized as
\begin{align}
	\frac{\rho_A}{s} \simeq\begin{cases}
	\displaystyle 3.7\times 10^{-10}\,{\rm GeV} \left( \frac{R_A}{0.1} \right)
	 \left( \frac{m_A}{10^{-8}\,{\rm GeV}} \right)^{1/2} \left( \frac{10^{14}\,{\rm GeV}}{H_{\rm inf}} \right)^{2/3}
	 \left( \frac{T_{\rm R}}{10^6\,{\rm GeV}} \right)^{4/3} & {\rm for}~~m_A < \Gamma_\phi  \\
	\displaystyle 3.5\times 10^{-10}\,{\rm GeV} \left( \frac{R_A}{0.1} \right)
	 \left( \frac{m_A}{1\,{\rm GeV}} \right)^{2/3} \left( \frac{10^{14}\,{\rm GeV}}{H_{\rm inf}} \right)^{2/3}
	  \left( \frac{T_{\rm R}}{10\,{\rm GeV}} \right) & {\rm for}~~m_A > \Gamma_\phi 
	\end{cases}.
\end{align}
It is consistent with the observed DM abundance ($\simeq 4\times 10^{-10}\,$GeV in terms of the energy to entropy density ratio) for wide parameter ranges.

\subsubsection{$\gamma >2$}

For $\gamma>2$, the vector energy density at the end of inflation is bounded as (\ref{rhoA_bound}). After inflation, the vector energy density scales as $\rho_A \simeq m_A^2 {\overline A}^2/2 \propto a^{-2}$. For $h=1$, for example, assuming the limit of instant reheating, i.e., the radiation dominated universe starts just after inflation, the final vector boson abundance is evaluated as
\begin{align}
	\frac{\rho_A}{s}\ll 10^{-26\gamma} \frac{H_{\rm inf} M_P^{1/2}}{m_A^{1/2}}
	\lesssim 10^{-13}\,{\rm GeV}\left( \frac{10^{-22}\,{\rm eV}}{m_A} \right)^{1/2}
	\left( \frac{H_{\rm inf}}{10^{14}\,{\rm GeV}} \right),
\end{align}
where we have taken $\gamma=2$ when evaluating the most right hand side. If the reheating is delayed, there is a further suppression factor of $(\Gamma_\phi/H_{\rm inf})^{1/3}$. Since the DM should be heavier than $\sim 10^{-22}\,$eV from the galactic structure~\cite{Hu:2000ke}, we conclude that it cannot explain total DM abundance as far as the standard thermal history is assumed in the early universe. It may be possible that the universe enters the kination regime before the completion of the reheating. In such a case the DM abundance can be enhanced, but we do not pursue this possibility in this paper.

For another choice $h(\phi)=f(\phi)$, as mentioned in Sec.~\ref{sec:inf2}, there is no strong suppression for the vector energy density at the end of inflation. The physical field $\overline A$ increases as $\overline A \propto a^{(\gamma-2)/2}$ during inflation and one can take $\overline A_{\rm end} \equiv \overline A(\tau_{\rm end})$ as a free parameter in this case. Taking the scaling $\rho_A \propto a^{-2}$ after inflation until $H\sim m_A$, as mentioned above and actually seen in Fig.~\ref{fig:25}, the final abundance is
\begin{align}
	\frac{\rho_A}{s} \simeq \frac{1}{8}\left( \frac{90}{\pi^2 g_*} \right)^{1/4}\left( \frac{\overline A_{\rm end}}{M_P} \right)^2
	\left( \frac{\Gamma_\phi}{H_{\rm inf}} \right)^{4/3}
	\left( \frac{m_A}{\Gamma_\phi} \right)\sqrt{m_A M_P},
\end{align}
for $m_A < \Gamma_\phi$, and
\begin{align}
	\frac{\rho_A}{s} \simeq \frac{T_{\rm R}}{8}\left( \frac{\overline A_{\rm end}}{M_P} \right)^2
	\left( \frac{m_A}{H_{\rm inf}} \right)^{4/3},
\end{align}
for $m_A > \Gamma_\phi$. Numerically we have
\begin{align}
	\frac{\rho_A}{s} \simeq\begin{cases}
	\displaystyle 2.5\times 10^{-11}\,{\rm GeV} \left( \frac{T_{\rm R}}{10^9\,{\rm GeV}} \right)^{2/3}
	 \left( \frac{m_A}{1\,{\rm GeV}} \right)^{3/2} \left( \frac{10^{14}\,{\rm GeV}}{H_{\rm inf}} \right)^{4/3}
	\left( \frac{\overline A_{\rm end}}{M_P} \right)^2 
	& {\rm for}~~m_A < \Gamma_\phi \\
	\displaystyle 1.3\times 10^{-10}\,{\rm GeV} \left( \frac{T_{\rm R}}{10^7\,{\rm GeV}} \right)
	 \left( \frac{m_A}{10^{2}\,{\rm GeV}} \right)^{4/3} \left( \frac{10^{14}\,{\rm GeV}}{H_{\rm inf}} \right)^{4/3}
	\left( \frac{\overline A_{\rm end}}{M_P} \right)^2 
	& {\rm for}~~m_A > \Gamma_\phi  
	\end{cases}.
\end{align}
It is possible to have a correct vector DM abundance in this case.

\section{Constraint from curvature and isocurvature perturbation} \label{sec:pert}

We have considered the dynamics of homogeneous mode of the vector boson in the previous section. 
Let us consider fluctuation of the inflaton and vector boson generated during inflation and its observational consequences. 

As shown in the previous section, when the vector energy density is negligible, the standard slow-roll inflation driven by just an inflaton field happens. We call this as ``isotropic'' regime. Then the vector boson backreaction to the inflaton becomes important if $\gamma<-4$ and the anisotropic inflation regime follows. The e-folding number of the anisotropic regime, $N_{\rm ani}$, depends on the initial condition.
If $N_{\rm ani} \gtrsim 60$ fluctuations of all the observable scale must arise during the anisotropic inflation, while if $N_{\rm ani} \lesssim 60$ the large scale fluctuations in the present universe may arise from the isotropic regime and only the small scale fluctuations may be affected by the anisotropic inflation. Thus we consider three cases for $\gamma<-4$: 
\begin{itemize}
\item (i) There is no anisotropic inflation regime $(N_{\rm ani}=0)$.
\item (ii) Anisotropic inflation regime is not long enough $(0< N_{\rm ani} \lesssim 60)$.
\item (iii) Anisotropic inflation regime lasts long enough $(N_{\rm ani}\gtrsim 60)$.
\end{itemize}
We note that there is only the case (i) for $\gamma>2$. We also assume that $h=1$ for $\gamma<-4$ and $h=f$ for $\gamma>2$.

\subsection{Isocurvature perturbation} \label{sec:iso}

It is convenient to move to the Fourier space and decompose the vector fluctuation into the transverse and longitudinal mode:
\begin{align}
	\delta \vec{A}(\vec x) = \int \frac{d^3k}{(2\pi)^3} \delta\vec{A}({\vec k}) e^{i\vec k\cdot \vec x}
	= \int \frac{d^3k}{(2\pi)^3} \left( \vec{A}_T(\vec k) + \hat k A_L(\vec k) \right)e^{i\vec k\cdot \vec x}.
\end{align}
where the transverse mode satisfies $\vec k\cdot \vec A_T=0$ and $\hat k\equiv \vec k/|\vec k|$. The action, in terms of $\vec A^f_T\equiv f\vec A_T$ and $A_L^g\equiv gA_L$ with $g\equiv f\sqrt{a^2h^2m_A^2/(a^2h^2m_A^2+f^2k^2)}$, is written as
\begin{align}
	&S = S_T + S_L,\\
	&S_T = \int \frac{d^3k d\tau}{(2\pi)^3} \frac{1}{2}\left[ \left| \vec{A}_T^{f\prime}(k)\right|^2 -\left(k^2+ \frac{a^2h^2m_A^2}{f^2}-\frac{f''}{f} \right)\left|\vec A_T^f(k) \right|^2  \right],  \label{ST}\\
	&S_L = \int \frac{d^3k d\tau}{(2\pi)^3} \frac{1}{2}\left[ \left| {A}_L^{g\prime}(k)\right|^2 -\left(k^2+ \frac{a^2h^2m_A^2}{f^2}-\frac{g''}{g} \right)\left|A_L^g(k) \right|^2  \right].  \label{SL}
\end{align}

The vector boson power spectrum is expressed as
\begin{align}
	\left< \delta A_{i}(\vec k) \delta A^{*}_j(\vec k') \right> = \frac{2\pi^2a^2}{k^3}\left[\mathcal P_T(k) (\delta_{ij}-\hat k_i \hat k_j) + \mathcal P_L(k) \hat k_i \hat k_j \right] (2\pi)^3 \delta(\vec k- \vec k'),
\end{align}
where $\mathcal P_T(k)$ and $\mathcal P_L(k)$ are the transverse and longitudinal power spectrum, respectively, which we will evaluate below. They are quantized by using the creation and annihilation operator as
\begin{align}
	&\vec{A}^f_T(\vec k,\tau) = \sum_{\lambda=\pm}\left[\widetilde A^f_T(\vec k,\tau) \vec\epsilon_{\lambda} a^T_{\lambda,\vec k}+ \widetilde A_T^{f*}(\vec k,\tau) \vec{\epsilon_{\lambda}^*} a^{T\dagger}_{\lambda,-\vec k}\right],\\
	&A^g_L(\vec k,\tau) = \widetilde A_L^g(\vec k,\tau) a^L_{\vec k}+ \widetilde A_{L}^{g*}(\vec k,\tau) a^{L\dagger}_{-\vec k},
\end{align}
where the polarization vector satisfies $\vec{\epsilon_{\lambda}^*}\cdot \vec{\epsilon_{\lambda'}}=\delta_{\lambda\lambda'}$ and also $\sum_\lambda \epsilon^\lambda_i \epsilon^{\lambda*}_j=\delta_{ij}-\hat k_i\hat k_j$. The creation and annihilation operator satisfy the commutation relation $\left[a^T_{\lambda, \vec k},a^{T\dagger}_{\lambda',\vec k'}\right] = (2\pi)^3\delta_{\lambda\lambda'}\delta(\vec k-\vec k')$ and so on.
In order to evaluate $\mathcal P_T(k)$ and $\mathcal P_L(k)$, we must now the time evolution of transverse and longitudinal fluctuation throughout the history of the universe. Details are summarized in App.~\ref{sec:time}. A short conclusion is that the transverse fluctuation is dominant for $\alpha\leq-4$ and $h=1$ and the longitudinal one is dominant for $\alpha\geq 2$ and $h=f$. Below we study the isocurvatrue perturbation in each case.

\subsubsection{Transverse mode $(\gamma\leq-4)$} \label{sec:trans}

The equation of motion of the transverse mode during inflation is given by
\begin{align}
	\widetilde A_T^{f\prime\prime}(k) + \left( k^2 + \frac{a^2h^2 m_A^2}{f^2} -\frac{\alpha(2+\alpha)}{4}\mathcal H^2 \right)\widetilde A_T^f(k) =0.
\end{align}
where we substituted $f^2\propto a^{\alpha}$. As explained in Sec.~\ref{sec:inf}, for $\gamma<-4$, the standard slow-roll inflation happens when the vector boson energy density is negligible and in this regime we have $\alpha=\gamma$. However, the inflationary universe will eventually enter the regime of anisotropic inflation supported by the vector condensate and in this regime we have $\alpha=-4$ independently of the value of $\alpha$. Let us define $\tau_{\rm ani}$ as the conformal time when the anisotropic inflation regime starts. We have
\begin{align}
	\alpha =  \begin{cases}
		\gamma & {\rm for}~~\tau < \tau_{\rm ani}\\
		-4 & {\rm for}~~\tau > \tau_{\rm ani}
	\end{cases}.
\end{align}
Thus the property of fluctuations of the observable scale depends on whether the present cosmological scale, $k_0^{-1}$, is longer or shorter than $|\tau_{\rm ani}|$. 

Neglecting the mass term, i.e., assuming $hm_A/f \ll H_{\rm inf}$, the solution to this equation is given by
\begin{align}
	\widetilde A_{T}(k,\tau) = e^{\frac{i(2\nu+1)\pi}{4}}\frac{1}{\sqrt{2k}} \sqrt{\frac{-\pi k\tau}{2}} H_\nu^{(1)}(-k\tau),~~~~~~
	\nu \equiv \frac{|1+\alpha |}{2},
	\label{AT_BD}
\end{align}
where $H_\nu^{(1)}(x)$ is the Hankel function of the first kind. The limiting form in the subhorizon and superhorizon limit are given by
\begin{align}
	\widetilde A^f_{T}(k,\tau) \simeq \begin{cases}
		\displaystyle \frac{1}{\sqrt{2k}}e^{-ik\tau} & {\rm for}~~ k/a \gg H_{\rm inf}\\
		\displaystyle e^{\frac{i(2\nu-1)\pi}{4}}\frac{a H_{\rm inf}}{\sqrt 2 k^{3/2}}\frac{\Gamma(\nu)}{\Gamma(3/2)} \left( \frac{2}{-k\tau} \right)^{\nu-3/2} & {\rm for}~~k/a \ll H_{\rm inf}
	\end{cases}.
\end{align}
Actually we have chosen the overall coefficient so that the mode function coincides with the Minkowski form in the short wavelength limit. Thus it evolves as $a^{\nu-1/2}$ after the horizon exit. The ``physical'' field $\overline{A_{T}}=\widetilde A^f_{T}/a$ evolves as $a^{\nu-3/2}$.
It is the same scaling as that of the homogeneous mode $\overline A$ (\ref{barAscaling}), as expected.

The transverse power spectrum after inflation is defined as
\begin{align}
	\left<\vec A_T(k)\cdot \vec A^*_T(k') \right> = \frac{4\pi^2 a^2}{k^3} \mathcal P_T(k) (2\pi)^3 \delta(\vec k- \vec k').
\end{align}
Now we evaluate it at the end of inflation. The shape of the spectrum depends on the case (i)--(iii). First, for the case (i) all the cosmologically relevant scales correspond to the modes that exit the horizon during the standard slow-roll inflation. Thus
\begin{align}
	\mathcal P_T(k) =\left( \frac{H_{\rm inf}}{2\pi} \right)^2\left( \frac{\Gamma(\nu)}{\Gamma(3/2)} \right)^2 \left( \frac{2aH_{\rm inf}}{k} \right)^{2\nu-3}.   \label{PT1}
\end{align}
Therefore, for $\nu > 3/2$ ($\alpha<-4$), the spectrum is red tilted. For the case (ii), large scale fluctuations $k< k_{\rm ani}$ experience both the standard slow-roll inflation and anisotropic inflation regime. Thus
\begin{align}
	\mathcal P_T(k) =\begin{cases}
	\displaystyle \left( \frac{H_{\rm inf}}{2\pi} \right)^2\left( \frac{\Gamma(\nu)}{\Gamma(3/2)} \right)^2 \left( \frac{2a_{\rm ani}H_{\rm inf}}{k} \right)^{2\nu-3} & {\rm for}~~k < k_{\rm ani} \\
	\displaystyle  \left( \frac{H_{\rm inf}}{2\pi} \right)^2 & {\rm for}~~k > k_{\rm ani}
	\end{cases}.
	\label{PT2}
\end{align}
For the case (iii), all the cosmologically relevant scales correspond to the modes that exit the horizon during the anisotropic inflation at which $\nu=3/2$, hence
\begin{align}
	\mathcal P_T(k) = \left( \frac{H_{\rm inf}}{2\pi} \right)^2.
	\label{PT3}
\end{align}
The typical magnitude of the fluctuation with a comoving wavenumber $k$ is given by $\sqrt{\mathcal P_T(k)}$. The isocurvature perturbation of the vector field is then given by
\begin{align}
	\sqrt{\mathcal P_S(k)} = \left.\frac{\delta \rho_A(k)}{\rho_A}\right|_{H=m_A} \simeq \left.\frac{2\sqrt{\mathcal P_T(k)}}{\overline A}\right|_{a=a_{\rm end}}
	\simeq \frac{H_{\rm inf}}{\pi \overline A_i},
	\label{SA_gamma-4}
\end{align}
where in the most right hand side we have defined $\overline A_i$ through $\overline A_i = \overline A_{\rm end} e^{-N_{\rm st}(\alpha+4)/2}$.
Here $N_{\rm st}$ denotes the e-folding number of the standard slow-roll inflation in the last 60 e-foldings, which satisfies $N_{\rm st}+N_{\rm ani}\simeq 60$. For the case (i) we have $N_{\rm st} \simeq 60$ while for the case (iii) we have $N_{\rm st}=0$ and the case (ii) lies between these two.
Since the time evolution of the transverse mode in the superhorizon limit is the same as the homogeneous one the final isocurvature perturbation can be evaluated at $a=a_{\rm end}$. The observational constraint is $\sqrt{\mathcal P_S(k)} \lesssim 9\times 10^{-6}$  at the present cosmological scale~\cite{Akrami:2018odb}.

\subsubsection{Longitudinal mode $(\gamma\geq 2)$} \label{sec:longi}

The evolution of longitudinal fluctuation for $H > hm_A/f$ is nontrivial in contrast to the transverse one. It is summarized in App.~\ref{sec:time}. We should evaluate $\widetilde A_L(k)$ at $H=m_A$ (note that $h=f=1$ after inflation), after which the superhorizon evolution becomes the same as the homogeneous mode.

We focus on the case of $\gamma\geq 2$ and $h=f$ since only in this case the longitudinal fluctuation is relevant, as shown in App.~\ref{sec:time}. From Eq.~(\ref{AL_after}), the longitudinal power spectrum at $H=m_A$ or $a=a_*$ is given by
\begin{align}
	\mathcal P_L(k) \simeq \left( \frac{H_{\rm inf}}{2\pi} \right)^2 \left( \frac{k}{a_* h_k m_A} \right)^2
	\simeq  \left( \frac{H_{\rm inf}}{2\pi} \right)^2  \left( \frac{k}{a_*m_A} \right)^2 \left( \frac{a_{\rm end}H_{\rm inf}}{k} \right)^\alpha. 
\end{align}
On the other hand, the homogeneous mode $\overline A$ at $H=m_A$ is evaluated as
\begin{align}
	\overline A(H=m_A) \simeq \overline A_{\rm end} \left( \frac{a_{\rm end}}{a_*} \right)
	\simeq \overline A_i \left( \frac{a_{\rm end}}{a_i} \right)^{\frac{\alpha-2}{2}} \left( \frac{a_{\rm end}}{a_*} \right),
\end{align}
where $a_i$ is the scale factor at the initial time, which should be at least 60 e-foldings before the end of inflation. Thus the isocurvature perturbation is
\begin{align}
	\sqrt{\mathcal P_S(k)} \simeq \left.\frac{2\sqrt{\mathcal P_L(k)}}{\overline A}\right|_{H=m_A} 
	\simeq \frac{H_{\rm inf}}{\pi \overline A_i} \left(\frac{k}{a_i m_A}\right) \left( \frac{a_i H_{\rm inf}}{k} \right)^{\alpha/2}
	\lesssim \frac{H_{\rm inf}}{\pi \overline A_i} \frac{H_{\rm inf}}{m_A},
	\label{SA_gamma2}
\end{align}
where in the last inequality we used $a_i < a_k = k/H_{\rm inf}$.


\subsection{Curvature perturbation} \label{sec:curv}

Here we briefly describe the statistical anisotropy in the curvature perturbation. Since there is a vector background during inflation, it can affect the statistical properties of the curvature perturbation. In particular, the power spectrum of the curvature perturbation may have the following quadrupolar asymmetric form:
\begin{align}
	\mathcal P_\zeta(\vec k) = \mathcal P^0_\zeta(k) \left[1+g_* \sin^2\theta_k\right],
	\label{Pzeta_aniso}
\end{align}
where $\mathcal P^0_\zeta(k)$ is the isotropic part of the dimensionless curvature perturbation power spectrum, normalized as $\mathcal P^0_\zeta(k) \simeq 2.1\times 10^{-9}$ at the present horizon scale~\cite{Akrami:2018odb}, $\theta_k$ is the angle between the wave vector $\vec k$ and the preferred direction and $g_*$ represents the magnitude of the statistical anisotropy. The observational constraint on this type of quadrupolar asymmetry is $|g_*| \lesssim 10^{-2}$~\cite{Akrami:2018odb}.
There are several effects that generates nonzero $g_*$ as extensively studied in e.g. Refs.~\cite{Maleknejad:2012fw,Bartolo:2012sd}. The dominant effect comes from the inflaton-vector boson interaction in the Lagrangian after expanding $f\simeq f_\phi \delta \phi$ and $\vec A = \vec A_0 + \vec{\delta A}$ around the homogeneous background, which gives the additional contribution to the inflaton 2-point function~\cite{Maleknejad:2012fw,Bartolo:2012sd}. 

Neglecting the interaction with the vector boson, the zeroth order solution for $\zeta$ during inflation is
\begin{align}
	&\zeta^0(\vec k,\tau) = \widetilde \zeta_{\vec k}(\tau) a_{\vec k}+ \widetilde \zeta^*_{\vec k}(\tau) a^\dagger_{-\vec k},\\
	&\widetilde \zeta_{\vec k}(\tau) = \frac{H_{\rm inf}}{\sqrt{2\epsilon}M_P} \frac{1+ik\tau}{\sqrt{2k^3}}e^{-ik\tau},~~~~~~
	\left[ a_{\vec k}, a^\dagger_{\vec k'} \right] = (2\pi)^3\delta(\vec k-\vec k').  
	\label{zetak}
\end{align}
Note that $\widetilde\zeta_{\vec k}=\widetilde\zeta_{-\vec k}$. The power spectrum at the zeroth order is given by
\begin{align}
	\left<\zeta^0(\vec k,\tau)\zeta^{0*}(\vec k',\tau) \right>= \frac{1}{2\epsilon M_P^2} \frac{H_{\rm inf}^2}{2k^3} (2\pi)^3 \delta(\vec k-\vec k')
	=\frac{2\pi^2}{k^3}\mathcal P^0_\zeta(k) (2\pi)^3 \delta(\vec k-\vec k'),
\end{align}
where $\mathcal P^0_\zeta =H_{\rm inf}^2/(8\pi^2\epsilon M_P^2)$.\footnote{
	In the anisotropic inflation regime, this $\epsilon$ should be regarded as $\epsilon = 16\epsilon_V/\gamma^2$.
} The unequal time correlator is given by
\begin{align}
	\left[\zeta^0(\vec k,\tau) ,\zeta^{0}(\vec k',\tau')  \right] = -\frac{i H_{\rm inf}^2}{6\epsilon M_P^2}(\tau^3-\tau^{'3}) (2\pi)^3 \delta(\vec k+\vec k').
	\label{zeta_unequal}
\end{align}
for the superhorizon limit $-k\tau \to 0$.

In the spatially flat gauge, the curvature perturbation is given by $\zeta=-H \delta\phi/\dot\phi = \delta\phi/(\sqrt{2\epsilon} M_P)$. The interaction action is then obtained as
\begin{align}
	S_{\rm int} = \int \frac{d\tau d^3k}{(2\pi)^3}\left[ -\gamma E^f_x \delta E^f_x(\vec k) \zeta(-\vec k) 
	- \gamma_h a^2 m_A^2 A^f_x \delta A^f_x(\vec k) \zeta(-\vec k)\right]
	\equiv -\int d\tau\,\mathcal H_{\rm int},
	\label{S_int}
\end{align}
where we defined $\gamma_h=0$ for $h=1$ and $\gamma_h=\gamma$ for $h=f$, $\vec E^f=-f\vec A'$ and we have taken $\vec A_0=(A_x,0,0)$ without loss of generality. Due to this interaction term, anisotropic curvature perturbation power spectrum appears at the level of second order perturbation in the Hamiltonian $\mathcal H_{\rm int}$.
We use the in-in formalism to calculate the two point function $\left<\zeta(\vec k,\tau) \zeta(\vec k',\tau)\right>$ at the end of inflation $\tau=\tau_{\rm end}$:
\begin{align}
	\left<\zeta(\vec k,\tau) \zeta(\vec k',\tau)\right>_{\rm 2nd}= -\int_{\tau_i}^{\tau} d\tau_1 \int_{\tau_i}^{\tau} d\tau_2 
	\left< \left[ \left[ \zeta^0_k(\tau)\zeta_{k'}^0(\tau), \mathcal H_{\rm int}(\tau_1)\right], \mathcal H_{\rm int}(\tau_2) \right] \right>
\end{align}
For $\gamma\leq -4$ and $h=1$ the first term in (\ref{S_int}) is important to evaluate the two point function, while for $\gamma\geq 2$ and $h=f$ the first term is irrelevant since $\vec E^f=0$ and the second term becomes important.

\subsubsection{$\gamma\leq -4$}

First let us consider the case of $\gamma\leq -4$ and $h=1$. After some computation, using the correlator (\ref{zeta_unequal}), we find
\begin{align}
	\left<\zeta(\vec k,\tau) \zeta(\vec k',\tau)\right>_{\rm 2nd}
	\simeq &\left( \frac{\gamma (\alpha+1)H_{\rm inf}^2}{6\epsilon M_P^2} \right)^2  \sin^2\theta_k (2\pi)^3\delta(\vec k+\vec k') \nonumber\\
	&\times \int_{\tau_i}^{\tau} d\tau_1 \int_{\tau_i}^{\tau} d\tau_2\,\frac{\tau^3-\tau_1^3}{\tau_1}\frac{\tau^3-\tau_2^3}{\tau_2}E_x^f(\tau_1)E_x^f(\tau_2) \widetilde A^f_T(k,\tau_1) \widetilde A^f_T(k',\tau_2).
\end{align}
where we used $1-\hat k_x^2 = \sin^2\theta_k$ with $\theta_k$ being the angle between the background vector field direction $\vec A$ and the wave vector $\vec k$. Note that $\delta \vec E^f = f\vec A_T' \simeq (\alpha+1)\vec A_T^f/\tau$ since $A_L'\simeq 0$ as explained in App.~\ref{sec:time}. Thus it gives the anisotropic power spectrum of the form (\ref{Pzeta_aniso}).
By using the solution (\ref{AT_BD}), the statistical anisotropy parameter $g_*$ is calculated as  
\begin{align}
	g_* \simeq \frac{\gamma^2(\alpha+1)^4}{18\epsilon M_P^2}\left( \frac{\Gamma(\nu)}{\Gamma(3/2)} \right)^2 \left(H_{\rm inf}A^f(\tau)\right)^2 \frac{k^{4+\alpha}}{(-2\tau)^{2+\alpha}} I(\tau),
\end{align}
where
\begin{align}
	I(\tau) \equiv \left[\int_{\tau_i}^\tau d\tau_1\, \tau_1^{\alpha+3}\right]^2=
	\begin{cases}
		\displaystyle \log^2(\tau/\tau_i) & {\rm for}~~\alpha=-4 \\
		\displaystyle \left(\frac{\tau^{\alpha+4}}{\alpha+4}\right)^2 & {\rm for}~~\alpha<-4
	\end{cases}.
\end{align}
Thus, for a particular case of $\alpha=-4$, we have
\begin{align}
	|g_*| = \frac{48 \rho_A(\tau)}{\epsilon\rho_\phi(\tau)} \log^2\left( \frac{\tau}{\tau_i} \right)
	~~\gtrsim~~ \frac{48 \rho_A(\tau)}{\epsilon\rho_\phi(\tau)} N^2(k),
\end{align}
which reproduces the result of Ref.~\cite{Bartolo:2012sd} where $N(k)$ represents the e-folding number of the inflation after the observable scale $k$ exit the horizon. For $\alpha<-4$, we have
\begin{align}
	|g_*| = \frac{\gamma^2(\alpha+1)^2}{3\cdot 2^{\alpha+4}(\alpha+4)^2}\left( \frac{\Gamma(\nu)}{\Gamma(3/2)} \right)^2
	\frac{\rho_A(\tau)}{\epsilon \rho_\phi(\tau)} (k\tau)^{\alpha+4}
	~~\gtrsim~~\mathcal C\frac{\rho_A(\tau)}{\epsilon \rho_\phi(\tau)} e^{-(\alpha+4)N(k)},
	\label{g*_gamma-4}
\end{align}
where $\mathcal C$ collectively represents a numerical factor, which is at least $\mathcal O(10)$ for reasonable value of $\gamma$.
The energy density is given as $\rho_A(\tau) \simeq (\alpha+1)^2(H_{\rm inf} A^f)^2/(2a^2).$
Recall that the value of $\alpha$ can change from $\gamma$ to $-4$ when the anisotropic inflation happens (case (ii) described in the beginning of this section). In such a case $N(k)$ should be regarded as an e-folding number of the standard slow-roll inflation, $N_{\rm st}\simeq 60-N_{\rm ani}$.

\subsubsection{$\gamma\geq 2$}

Next we consider the case of $\gamma\geq 2$ and $h=f$. As already mentioned, the second term of (\ref{S_int}) gives dominant contribution to the anisotropic power spectrum. The computation is parallel to the previous case of $\gamma \leq -4$, except that the longitudinal fluctuation gives the dominant contribution in this case.\footnote{
	Thus the power spectrum has of the dependence of $(\hat k_x)^2=\cos^2\theta_k = 1-\sin^2\theta_k$ and hence the sign of $g_*$ may be flipped in this case.
}
Noting $A_L^f(\tau) \sim (H_{\rm inf}/m_A) A_T^f(\tau)$ at $\tau=\tau_{\rm end}$, the result is
\begin{align}
	|g_*| \sim \begin{cases}
		\displaystyle \frac{48\rho_A(\tau)}{\epsilon \rho_\phi(\tau)} N^2(k) & {\rm for}~~\alpha=2 \\
		\displaystyle \mathcal C\frac{\rho_A(\tau)}{\epsilon \rho_\phi(\tau)} e^{(\alpha-2)N(k)} & {\rm for}~~\alpha>2
	\end{cases}.
	\label{g*_gamma2}
\end{align}
Note that $\rho_A(\tau) \simeq (m_A A^f)^2/(2a^2)$ in this case. Thus the statistical anisotropy can be suppressed compared with the previous case.

\subsection{Constraint} 

Now we are going to discuss constraint on the vector DM scenario from the isocurvature perturbation. First we consider $\gamma <-4$. 
In this case the isocurvature perturbation is given by (\ref{SA_gamma-4}) while the statistical anisotropy parameter is given by (\ref{g*_gamma-4}).
Combining them we obtain
\begin{align}
	|g_*| \sim \mathcal C \frac{\mathcal P_\zeta^0}{\mathcal P_S}e^{-2(\alpha+4)N_{\rm st}}.  \label{g*_case1}
\end{align}
Taking account of the constraint on the isocurvature perturbation, $|g_*|$ must be much larger than one and it clearly contradicts with the observational constraint. For $\gamma=-4$, the factor $e^{-2(\alpha+4)N_{\rm st}}$ should be replaced with $(N_{\rm st}+N_{\rm ani})^2\sim 3600$ and this case is also excluded.\footnote{
	In Ref.~\cite{Nakayama:2019rhg} constraint from the statistical anisotropy of the curvature perturbation was not taken into account.
}

For $\gamma>2$, an important difference from the $\gamma<-4$ case is that the energy density of the vector homogeneous background can be extremely small: $\rho_A \simeq m_A^2 \overline A^2/2$ for $h=f$, since the kinetic energy vanishes for the $C_1$ solution in (\ref{A_inf}). Therefore, the statistical anisotropy (\ref{g*_gamma2}) can be suppressed.
On the other hand, the longitudinal fluctuation contributes to the isocurvature perturbation (\ref{SA_gamma2}) and it gives stringent constraint on this scenario. 
The statistical anisotropy parameter (\ref{g*_gamma2}) is rewritten as
\begin{align}
	|g_*| \sim \mathcal C \frac{\mathcal P_\zeta^0}{\mathcal P_S} e^{2(\alpha-2)N(k)}.   \label{g*_case2}
\end{align}
The expression is similar to the previous case and it is clearly too large once we impose the isocurvature constraint. For $\gamma=2$, the factor $e^{2(\alpha-2)N(k)}$ should be replaced with $N^2(k)\sim 3600$ and this case is also excluded.
The reason that these two constraints (isocurvature and statistically anisotropic curvature perturbation) are complementary is that the inflaton-vector coupling is enhanced by $1/\epsilon$ in order to realize the scaling $f\propto a^{\alpha/2}$. To suppress the isocurvature perturbation one requires small $H_{\rm inf}$, which makes the $1/\epsilon$ coupling stronger and the statistical anisotropy becomes larger.

A loophole is that the inflaton may not responsible for the observed curvature perturbation. In the curvaton scenario, the observed curvature perturbation is originated from the other scalar field fluctuation than the inflaton, called the curvaton~\cite{Enqvist:2001zp,Lyth:2001nq,Moroi:2001ct}. In this case $\mathcal P_\zeta^0$ appearing e.g. in Eq.~(\ref{g*_case1}) or (\ref{g*_case2}) should be interpreted as the sub-dominant inflaton contribution to the curvature perturbation, and it can take much smaller value than the observed value. In such a case the constraint from the statistical anisotropy may be avoided.

\section{Conclusions and discussion} \label{sec:dis}

In this paper we studied scenario for vector coherent oscillation DM with the action given by (\ref{action}). The homogeneous vector condensate can be formed for $\gamma\leq -4$ or $\gamma \geq 2$. 
The particular case of $\gamma=-4$ and $2$ was studied in Ref.~\cite{Nakayama:2019rhg} and in this paper we mainly considered $\gamma<-4$ and $\gamma>2$.

For $\gamma<-4$, the vector condensate energy density increases during inflation and eventually the backreaction becomes important and the so-called anisotropic inflation occurs~\cite{Watanabe:2009ct,Soda:2012zm,Maleknejad:2012fw}. It is indeed possible that the vector condensate will become a coherent oscillation and its abundance is consistent with the observed DM abundance. However, it is found that the combination of constraints from DM isocurvature fluctuation and also the statistical anisotropy of the curvature perturbation almost exclude vector coherent oscillation DM scenario. For $\gamma>2$, the vector abundance crucially depends on the form of the mass function $h$ in (\ref{action}). For the simplest case $h=1$ the vector coherent oscillation abundance is too low to explain total DM. For $h=f$, the vector coherent oscillation can be total DM. However, the combination of constraints from DM isocurvature fluctuation and the statistical anisotropy of the curvature perturbation also exclude this scenario.

A possible loophole is that the inflaton is not responsible for the observed curvature perturbation and the curvaton explains the curvature perturbation, or thermal history after inflation has an epoch of non-standard equation of state such as kination regime.
As another possibility we may consider the Higgs mechanism instead of the Stuckelberg mechanism to generate the vector boson mass $m_A$. It is possible that the Higgs is stabilized at the symmetric phase during inflation so that there is no development of longitudinal fluctuation and the symmetry breaking occurs after inflation. For $\gamma\geq 2$, the problematic isocurvature perturbation may be avoided in such a case, although there are extra contributions to the vector abundance from the Higgs decay or the cosmic string dynamics.

To summarize, the vector coherent oscillation DM is severely restricted from cosmological observation and some additional modifications are required to make this scenario viable. If it constitutes the present DM, it may be detectable by experiments proposed so far~\cite{Horns:2012jf,Parker:2013fxa,Chaudhuri:2014dla,Hochberg:2016ajh,Hochberg:2016sqx,Bloch:2016sjj,Hochberg:2017wce,Arvanitaki:2017nhi,Knapen:2017ekk,Baryakhtar:2018doz,Griffin:2018bjn,Chigusa:2020gfs} through the (small) kinetic mixing between the vector boson and the Standard Model photon.

\section*{Acknowledgments}

This work was supported by the Grant-in-Aid for Scientific Research C (No.18K03609 [KN]) and Innovative Areas (No.17H06359 [KN]).

\appendix
\section{Time evolution}  \label{sec:time}

In this Appendix we summarize time evolution of the homogeneous mode, transverse mode and longitudinal mode.

\subsection{Homogeneous mode}

The homogeneous equation is
\begin{align}
	0&= f^2 \vec A'' + 2ff' \vec A' + a^2 h^2 m_A^2 \vec A \\
	&= \vec A^{f\prime\prime} + \left( \frac{a^2h^2m_A^2}{f^2} - \frac{f''}{f} \right)\vec A^f,
\end{align}
where $A^f\equiv fA$. Note that
\begin{align}
	\frac{f''}{f} =\frac{\alpha(\alpha-1-3w)}{4} \mathcal H^2,
\end{align}
where $w$ is the equation-of-state parameter of the universe: $w=-1$ for the de Sitter universe, $w=0$ $(w=1/3)$ for the matter (radiation) dominated universe.

The solution during inflation is,
\begin{align}
	&A \propto c_1 +c_2 a^{-(\alpha+1)}, \\
	&A^f \propto c_1 a^{\alpha/2}+c_2 a^{-(\alpha+2)/2}, 
	&\overline A \propto c_1 a^{(\alpha-2)/2}+c_2 a^{-(\alpha+4)/2}.
\end{align}
where $\overline A= A^f/a$ is the ``physical'' vector field. The solution after inflation is
\begin{align}
	&A=A^f \propto d_1+ d_2\tau \propto d_1 +d_2 a^{(3w+1)/2},\\
	&\overline A \propto d_1 a^{-1}+d_2 a^{(3w-1)/2}.
\end{align}
For $\alpha\leq -4$ the $d_2$-term solution applies and for $\alpha\geq 2$ the $d_1$-term solution applies.

\subsection{Transverse mode}

The equation of the transverse mode, $\widetilde A_T^f \equiv f \widetilde A_T$, is
\begin{align}
	0 = \widetilde A^{f\prime\prime}_T(\vec k) + \left( k^2+\frac{a^2h^2m_A^2}{f^2} - \frac{f''}{f} \right)\widetilde A^f_T(\vec k).
\end{align}
It satisfies the same equation as the homogeneous mode $A^f$ for $k/a \ll H$.
Thus the solution during inflation is
\begin{align}
	&\widetilde A_T^f \propto c_1 a^{\alpha/2}+c_2 a^{-(\alpha+2)/2},\\
	&\widetilde A_T \propto c_1 +c_2 a^{-(\alpha+1)}.
\end{align}
The solution after inflation is
\begin{align}
	\widetilde A_T=\widetilde A_T^f \propto d_1+ d_2\tau \propto d_1 +d_2 a^{(3w+1)/2}.
\end{align}
As noted above, for $\alpha\leq -4$ the $d_2$-term solution applies and for $\alpha\geq 2$ the $d_1$-term solution applies.

By using these solutions and the Bunch-Davies initial condition (\ref{AT_BD}), $\widetilde A_T(k)$ at $H=m_A$ or $a=a_*$ is evaluated as follows. For $\alpha\leq-4$ and $h=1$, it is given by
\begin{align}
	\widetilde A_T(k,H=m_A) \sim \frac{1}{\sqrt{2k}}\left(\frac{a_{\rm end}}{a_k}\right)^{-\frac{\alpha+2}{2}}\left(\frac{a_*}{a_{\rm end}}\right)^{\frac{3w+1}{2}}
	= \frac{a_* H_{\rm inf}}{\sqrt{2}k^{3/2}} \left(\frac{a_{\rm end} H_{\rm inf}}{k}\right)^{-\frac{\alpha+4}{2}}\left(\frac{a_*}{a_{\rm end}}\right)^{\frac{3w-1}{2}},
	\label{AT_h=1}
\end{align}
where $a_k=k/H_{\rm inf}$ denotes the scale factor at the horizon exit during inflation.\footnote{
	Note that $\alpha$ as well as $w$ may not be constant: as explained in Sec.~\ref{sec:dyn}, the value of $\alpha$ can change when the anisotropic inflation happens. In the reheating phase $w$ also changes from some value (assumed to be $0$ in the most part of this paper) to $w=1/3$. Eq.~(\ref{AT_h=1}) should be understood as a shorthand notation and it implicitly includes such an effect.
	}
For $\alpha>2$ and $h=f$, on the other hand, $\widetilde A_T(k)$ (not $\widetilde A^f_T(k)$) remains constant in the superhorizon regime and hence it is evaluated at $a=a_k$, i.e., at the horizon exit during inflation. Thus we have
\begin{align}
	\widetilde A_T(k,H=m_A) \sim \frac{1}{\sqrt{2k}}\frac{1}{f(a=a_k)}=\frac{a_* H_{\rm inf}}{\sqrt{2} k^{3/2}} \left( \frac{a_{\rm end} H_{\rm inf}}{k} \right)^{\frac{\alpha-2}{2}} \left( \frac{a_{\rm end}}{a_*} \right).
	\label{AT_h=f}
\end{align}

\subsection{Longitudinal mode}

The equation of motion of the longitudinal mode $\widetilde A_L^g\equiv g\widetilde A_L$ is
\begin{align}
	0 = \widetilde A^{g\prime\prime}_L(\vec k) + \left( k^2+\frac{a^2h^2m_A^2}{f^2} - \frac{g''}{g} \right)\widetilde A^g_L(\vec k),
\end{align}
where
\begin{align}
	g= f\sqrt{\frac{a^2h^2m_A^2}{a^2h^2m_A^2+f^2k^2}}
	\simeq \begin{cases}
	ahm_A/k & {\rm for}~~fk\gg a hm_A \\
	f & {\rm for}~~fk\ll a hm_A
	\end{cases}
\end{align}

\subsubsection{During inflation}

For $h=1$, during inflation, we have
\begin{align}
	\frac{g''}{g} &= \frac{f''}{f} + \frac{2-\alpha}{4} \frac{f^2k^2\left[ (4+\alpha)f^2k^2+2(2\alpha-1)a^2m_A^2 \right]}{(f^2k^2+a^2m_A^2)^2} \mathcal H^2 \\
	&= \frac{8f^4k^4-2(\alpha^2-7\alpha+2)f^2k^2a^2m_A^2+\alpha(\alpha+2)a^4m_A^4 }{4(f^2k^2+a^2m_A^2)^2} \mathcal H^2\\
	&\simeq \begin{cases}
		\displaystyle 2\mathcal H^2 & {\rm for}~~fk\gg a m_A \\
		\displaystyle \frac{\alpha(\alpha+2)}{4}\mathcal H^2 & {\rm for}~~fk\ll a m_A
	\end{cases}.
\end{align}
Thus
\begin{align}
	\widetilde A_L^g \propto \begin{cases}
	c_1 a^{1}+c_2 a^{-2} & {\rm for}~~fk\gg a m_A \\
	c_1' a^{\alpha/2}+c_2' a^{-(\alpha+2)/2} & {\rm for}~~fk\ll a m_A
	\end{cases}
\end{align}
In term of the original basis, it is equivalent to
\begin{align}
	\widetilde A_L \propto \begin{cases}
	c_1 +c_2 a^{-3} & {\rm for}~~fk\gg a m_A \\
	c_1' +c_2' a^{-(\alpha+1)} & {\rm for}~~fk\ll a m_A
	\end{cases}
\end{align}
As noted in Ref.~\cite{Nakayama:2019rhg}, for $\alpha=-4$, the initial growing solution ($c_1$ term) connects to the final decaying solution ($c_1'$ term). We confirmed the same behavior numerically also for $\alpha<-4$.
Note that the initial evolution, $\widetilde A_L^g\propto a$, is slower than $\widetilde A_T^f$, for $\alpha<-4$ and $\alpha>2$. 

For $h=f$, during inflation, we have
\begin{align}
	\frac{g''}{g} 
	&=\frac{(\alpha+2)(\alpha+4)k^4+2(\alpha^2+4\alpha-2)k^2 a^2m_A^2 + \alpha(\alpha+2)a^4 m_A^4}{4(k^2+a^2m_A^2)^2}\mathcal H^2 \\
	&\simeq \begin{cases}
		\displaystyle \frac{(\alpha+2)(\alpha+4)}{4}\mathcal H^2 & {\rm for}~~k\gg a m_A \\
		\displaystyle \frac{\alpha(\alpha+2)}{4}\mathcal H^2 & {\rm for}~~k\ll a m_A
	\end{cases}
\end{align}
Thus
\begin{align}
	\widetilde A_L^g \propto \begin{cases}
	c_1 a^{(\alpha+2)/2}+c_2 a^{-(\alpha+4)/2} & {\rm for}~~k\gg a m_A \\
	c_1' a^{\alpha/2}+c_2' a^{-(\alpha+2)/2} & {\rm for}~~k\ll a m_A
	\end{cases}
\end{align}
In term of the original basis, it is equivalent to
\begin{align}
	\widetilde A_L \propto \begin{cases}
	c_1 +c_2 a^{-(\alpha+3)} & {\rm for}~~k\gg a m_A \\
	c_1' +c_2' a^{-(\alpha+1)} & {\rm for}~~k\ll a m_A
	\end{cases}.
\end{align}
Since we only consider $\alpha \geq 2$ for $h=f$, we have the initial $c_1$ solution and it connects to the $c_1'$ solution. We also numerically checked it. After all, $\widetilde A_L$ remains constant for all the cases of our interest.

\subsubsection{After inflation}

After inflation $f=h=1$ and
\begin{align}
	g = \sqrt{ \frac{a^2m_A^2}{k^2+a^2m_A^2}}
	\simeq \begin{cases}
	am_A/k & {\rm for}~~k\gg a m_A \\
	1 & {\rm for}~~k\ll a m_A
	\end{cases}
\end{align}
Thus
\begin{align}
	\frac{g''}{g}&= \frac{(1-3w)k^4-(5+3w)k^2 a^2m_A^2}{2(k^2+a^2m_A^2)^2}\mathcal H^2\\
	&\simeq \begin{cases}
		\displaystyle \frac{1-3w}{2}\mathcal H^2 =  \frac{a^2 R}{6}& {\rm for}~~k\gg a m_A \\
		\displaystyle -\frac{5+3w}{2}\frac{k^2}{a^2m_A^2}\mathcal H^2 & {\rm for}~~k\ll a m_A
	\end{cases}.
\end{align}
The solution is
\begin{align}
	\widetilde A_L^g \propto \begin{cases}
	d_1 a+d_2 a^{(3w-1)/2} & {\rm for}~~k\gg a m_A \\
	d_1'+d_2' a^{(1+3w)/2} & {\rm for}~~k\ll a m_A
	\end{cases}
\end{align}
Therefore $\widetilde A_L=\widetilde A_L^f = \widetilde A^g_L/g$ evolves as
\begin{align}
	\widetilde A_L=\widetilde A_L^f \propto \begin{cases}
	d_1+d_2 a^{3(w-1)/2} & {\rm for}~~k\gg a m_A \\
	d_1'+d_2' a^{(1+3w)/2} & {\rm for}~~k\ll a m_A
	\end{cases}
\end{align}
We numerically check that, starting from the $d_1$ solution (growing) solution as an initial condition, it connects to the $d_1'$ solution both for $w=0$ and $w=1/3$, even though apparently the $d_2'$ solution seems to become dominant.

Combining the solutions during and after inflation, we conclude that $\widetilde A_L(k)$ (not $\widetilde A_L^g(k)$) for the superhorizon mode remains constant until $H=m_A$ for all the cases of our interest. It means that $\widetilde A_L(k)$ at $H=m_A$ is the same as $\widetilde A_L(k)$ at $a=a_k=k/H_{\rm inf}$, i.e., at the horizon exit during inflation. Thus we have
\begin{align}
	\widetilde A_L(k,H=m_A) \sim \frac{1}{\sqrt{2k}}\frac{1}{g(a=a_k)} \simeq 
	\begin{cases}
		\displaystyle\frac{a_* H_{\rm inf}}{\sqrt{2} k^{3/2}} \frac{k}{a_* m_A} & {\rm for}~~h=1\\
		\displaystyle\frac{a_* H_{\rm inf}}{\sqrt{2} k^{3/2}} \frac{k}{a_* h_k m_A} & {\rm for}~~h=f
	\end{cases},
	\label{AL_after}
\end{align}
where $h_k\equiv h(a=a_k)$. 
Comparing it with the transverse solution (\ref{AT_h=1}), the ratio is
\begin{align}
	\frac{\widetilde A_L(k,H=m_A)}{\widetilde A_T(k,H=m_A)} \simeq \left(\frac{k}{a_* m_A}\right)\left( \frac{k}{a_{\rm end} H_{\rm inf}} \right)^{-\frac{\alpha+4}{2}} \left( \frac{a_{\rm end}}{a_*} \right)^{\frac{3w-1}{2}},
\end{align}
for $\alpha\leq -4$ and $h=1$. Whether it is larger than unity or not depends on the precise value of $\alpha$ and the duration of reheating period, i.e., the duration of $w=0$. Practically, unless $\alpha$ is very close to $-4$ and the reheating temperature is very low, it is likely that this ratio is smaller than unity and hence the transverse fluctuation is dominant.
On the other hand, the ratio is evaluated as
\begin{align}
	\frac{\widetilde A_L(k,H=m_A)}{\widetilde A_T(k,H=m_A)} \simeq \frac{f(a=a_k)}{g(a=a_k)} = \frac{k}{a_k m_A} = \frac{H_{\rm inf}}{m_A},
\end{align}
for $\alpha\geq 2$ and $h=f$. Thus the longitudinal fluctuation is much larger than the transverse one in this case.



\end{document}